%% file: main.tex
\documentclass[12pt]{article}

\usepackage{coling2018}
\usepackage{url}
\usepackage[para,online,flushleft]{threeparttable}
\usepackage{booktabs}
\usepackage{float}
\usepackage{graphicx,subcaption}
\usepackage{colortbl}
\usepackage{amsmath}
\usepackage{array}
\usepackage{tabularx}
\usepackage{tikz}
\usetikzlibrary{arrows.meta, positioning, matrix}
\usepackage{pgfplots}
\pgfplotsset{compat=1.18}
\usepackage[stable]{footmisc}

\date{}
\definecolor{lightgray}{rgb}{0.95,0.95,0.95}
\begin{document}

\title{Voices of the Mountains: Deep Learning-Based Vocal Error Detection System for Kurdish Maqams}

\author{
	\begin{tabular}[t]{c}
		Darvan Shvan Khairaldeen and Hossein Hassani\\
		\textnormal{University of Kurdistan Hewl\^er}\\
		\textnormal{Kurdistan Region - Iraq}\\
		{\tt \{darvan.shfan, hosseinh\}@ukh.edu.krd}
	\end{tabular}
}

\maketitle

\begin{abstract}
Maqam, a singing type, is a significant component of Kurdish music. Maqam-bezh (maqam-singer) receives training in a traditional face-to-face or through self-training. Automatic Singing Assessment (ASA) uses machine learning (ML) in providing the accuracy of singing styles and can help learners to improve their performance through error detection. Currently, the available ASA tools follow Western music rules. The musical composition requires all notes to stay within their expected pitch range from start to finish. The system fails to detect micro-intervals and pitch bends, so it identifies Kurdish maqam singing as incorrect even though the singer performs according to traditional rules. For Kurdish maqam, the computational process requires recognizing performance errors within microtonal spaces, which is beyond Western equal temperament. This research is the first attempt to address the mentioned gap. While many error types happen during singing, this research focuses on pitch, rhythm, and modal-stability errors. To control the scope and extent of work, we focused on Bayati-Kurd and developed a detection-and-classification-based model using deep learning. We created a corpus of 50 songs (22{,}050 Hz) from 13 vocalists (~2--3 hours) and annotated 221 error spans (150 fine pitch, 46 rhythm, 25 modal drift), exposing a strong class imbalance. It was segmented into 15{,}199 overlapping windows (10 s / 1 s hop; 3 s / 0.5 s hop for augmentation) and converted to log-mel spectrograms. We developed a two-headed CNN--BiLSTM with attention mode to decide whether a window contains an error and to classify it based on the chosen errors. Trained for 20 epochs with early stopping at epoch 10, the model reached validation macro-F1 of 0.468 (detection F1 0.216). On the full 50-song evaluation at a 0.750 threshold, recall was 39.4\% and precision 25.8\% (F1 = 0.311). Within detected windows, type macro-F1 was 0.387, with F1 of 0.492 (fine pitch), 0.536 (rhythm), and 0.133 (modal drift); modal drift recall was 8.0\% (5/25 events, 2 correctly typed). The better performance on common error types shows that the method works, while the poor modal-drift recall shows that more data and balancing are needed. This research opens a new track in ML regarding one of the least-studied aspects of Kurdish culture. It helps researchers and developers use ML and artificial intelligence to preserve and expand one of the central aspects of Kurdish culture.
\end{abstract}

\section{Introduction}

Singers need feedback to identify their musical mistakes that affect their pitch, timing, and phrasing delivery. The teacher responsible for this duty assists students, but not every student receives this support. The combination of lesson expenses, time restrictions, and restricted educational opportunities hinders certain individuals from acquiring the necessary musical training. Artificial Intelligence (AI)-based singing assessment could assist in removing that hindrance. Older automated systems operated through the implementation of a rigid set of rules. They used dynamic time warping algorithms to study pitch curves, but they perform best when the singing closely follows a fixed reference melody \cite{gupta2022taslp}. They generated incorrect results when performers used expressive vocal methods and when they performed songs that differed from Western musical standards.

The current models learn through audio-based learning systems. Triplet networks serve as one example. The system employs feedback to improve excellent performance towards the reference standard while driving low-quality performance away from the reference \cite{zhang2021ismir}, consequently producing more stable feedback. The transformer model in Polytune performs two functions: converting audio data into musical notation and simultaneously identifying pitch and rhythm errors \cite{chou2025polytune}. Basic networks, which focus on pitch and tempo, have achieved better results than traditional tools developed based on manually crafted rules \cite{wang2022complexity}. Recently, researchers have applied these concepts to analyze various musical traditions. For example, some focused on training neural networks to detect Kurdish music genres by analyzing acoustic signals \cite{zuhair2021kurdish}. Research modeled microtonal pitch distributions directly from audio recordings rather than assuming equal-tempered pitch bins, according to their study \cite{gedik2010hist,bozkurt2014review}. The research between Kurdish and Iranian maqams showed shared structural components \cite{ziaoddini2020maqam}, but other studies analyzed vocal pitch contours and maqam structures using pitch-sensitive representations such as CQT and ornament analysis methods \cite{yesiler2018makam,shafiei2019ornamentation}.

The majority of Automatic Singing Assessment (ASA) tools operate based on established Western music standards. The musical tradition requires performers to sing only the twelve established notes while maintaining absolute pitch precision. The system identifies smooth pitch bends and quarter-tones that singers perform as incorrect notes. The system lacks the ability to change its musical preferences or modify its assessment criteria for different cultural backgrounds, so it identifies non-Western vocal performances as incorrect \cite{gupta2022taslp}.

The Kurdish musical tradition uses musical modes, which they refer to as Maqams. The musical elements of each maqam present themselves through its unique interval patterns, melodic structures, and phrasing methods. The Bayat and Hijaz families maintain their own musical heritage, but Sega, Chwarga, and Bayati-Kurd operate as separate musical genres. Musicians don't agree on how many maqams there are or what they're called. The style is taught by ear, so these differences are normal \cite{hashemi2021kurdishmaqams,ziaoddini2020maqam,nettl2005ethnomusicology}.

People learn to sing through the process of copying what other singers do. The students follow their teacher's singing while the teacher provides immediate correction. The training process exists as an inaccessible resource for everyone who wants to learn singing independently, because independent singing without feedback leads to developing singing habits that become difficult to fix during future sessions.

This research examines Bayati-Kurd to prepare a model that detects three musical mistakes that affect \textit{pitch accuracy}, \textit{rhythm precision}, and \textit{modal pitch stability}. We create a dataset, which Kurdish music experts annotate by identifying errors, marking both the start and end positions that contain the mistakes. We develop a model to examine individual segments of the singer's performance in sequential order. The assessment evaluates three essential elements: \textit{pitch}, \textit{rhythm}, and \textit{modal drift}. The model's feedback shows both incorrect actions and their exact occurrence times, and produces results that experts use to verify a song against professional bases, assisting users during their professional activities. It can be incorporated into tools to serve as a teaching instrument to help singers who do not receive regular music education gain extra guidance. 

The rest of this paper is organized into four sections: related work, method, results and discussion, and conclusion.

\section{Related Work}

Research on automatic singing evaluation has developed from initial alignment and rule systems to learning-based models that can function with larger and noisier datasets. Most of these works are based on Western tonal music with equal-tempered tuning as well as static pitch targets. The following sections review some of the essential works related to the research topic. 

\subsection{Automatic Singing Assessment}

The early 2000s saw the start of work in Automatic singing assessment ASA systems aimed to evaluate the singing voice of a particular individual and offer ways to improve it. The first systems developed in the ASA domain were focused on pitch and time and used rudimentary signal processing and pitch detection frameworks. It aimed at comparing the singing to a target score/recorded reference. Most recently, these systems have employed deep learning approaches. They are now capable of analyzing audio and spectrograms to a greater variety of quality and recording conditions. While they are based on equal-tempered tuning, fixed pitch targets, and scoring rules congruent with Western teaching, Kurdish maqam singing employs microtonal intervals and expressive pitch movement, which do not fit these beliefs.

First, ASA systems were custom-built systems that utilized rules for comparing learner output with exemplar output. Dynamic Time Warping (DTW) was often utilized to match an executed pitch contour to a target pitch contour, even when there was a difference in tempo. DTW alignment methods have limitations when dealing with expressive timing and ornamentation. \cite{gupta2022taslp}. DTW finds an alignment that brings two sequences closer together. It can handle tempo differences, but it still sees many expressive pitch changes as mismatches when the reference has fixed pitch targets. DTW-based models highlighted an inherent problem present during the initial stages of ASA design. The objective of alignment was to incentivize the user to get as close as possible to the target contour. There are instances when an expressive rendition of a song may include a stretched and/or ornamented rhythm, as well as technically appropriate fluid pitch changes. Sometimes, such instances may deviate from the target contour, resulting in alignment errors. However, such discrepancies should not be highlighted as errors, as they are musically and stylistically appropriate.

Other early works focused on using modified acoustic information and basic predictors. Some systems tried to combine measures related to pitch with energy and/or spectral features and then trained regressors to model perceived quality. For instance, \newcite{wang2022complexity} studied the relationship between engineered acoustic features and perceived singing quality using a learned predictor. The functionality of these models depends on the definition of features and the style of the training data. A shift in repertoire, tuning, or vocal style can alter performance. However, they can provide feedback on a limited basis, concerning pitch-mismatch locations, timing-mismatch locations, or global summary scores. The outputs are useful for novice practice on fixed melodies. They restrict assistance for traditions where the accuracy relies on the mode behavior across phrases or where the pitch motion within a single note is musically significant.

Using statistical machine learning has brought a new type of ASA systems. Linear regression, SVMs, GMMs, and HMMs were used to score and classify musical attributes. While the ASA development-based training and using linear regression appeared to be straightforward, the assumption of a simple link between features and quality did not work for expressive singing in all cases. Although statistical models demonstrated greater flexibility, the main constraint remained the same, which was the handcrafted features and reflection of Western singing conditions with limited tuning variability. Systems based on statistical features of measurable audio parameters have tried to improve the interpretability of feedback \cite{gupta2022taslp,wang2022complexity}. Research has included comparing pitch range and variance, stability metrics, spectral tilting, and energy balance to expert evaluations. This complements diagnostic feedback, for example, for unstable intonation, or weak note onsets. This potential will run into challenges when singers employ stylistic approaches to singing that are not represented in the training data, and the same acoustic parameter in example traditions can have widely different meanings. A different problem that keeps coming up is cross-singer generalization. When models are trained on small groups of singers, they can learn the specific identities of the singers interspersed within the timbre or pitch range. This lessens the usefulness of the models in educational contexts where learners of different vocal ranges, timbres, dialects, and training backgrounds are present \cite{gupta2022taslp}.

In contrast with statistical machine learning,  deep learning models capture and learn the internal representation of audio according to spectrogram images. This assists in modeling pitch-related spectral salience patterns from audio recordings. This area also significantly helped to move away from the need to manually capture and describe each of the acoustic features. Deep learning models are highly dependent on the characteristics of the training dataset. If the training dataset captures only a limited range of styles, then the model representations will also be limited to that range \cite{gupta2022taslp}. This is particularly problematic for microtonal traditions, because the target pitches and the rules governing pitch changes are different from the Western training data. 

A CNN (Convolutional Neural Network) model can learn local time-frequency representations from spectrogram input \cite{hershey2017cnn,choi2017attention}. Recently, CNNs have been applied in melody extraction and pitch-related tasks, capturing spectro-temporal patterns contained in the $f_0$ structure \cite{bittner2017deep}. This is relevant because ASA systems require reliable pitch representations extracted from audio recordings.

LSTM (Long-Short Term Memory) and BiLSTM (Bidirectional LSTM) are two types of RNN (Recurrent Neural Network) that can model temporal structure and remember longer contexts. This ability is useful for things like modeling the start of a song, breaking down notes into smaller parts, and learning how to time patterns in sequences \cite{graves2013speech}.

The longer context is especially important for singing because the structure of a phrase can significantly affect accuracy. Hybrid setups are models such as CRNNs (Convolutional RNN) that use CNNs to pull out features and RNNs to model time. This architecture is widely used for audio applications that require looking at short spectral cues while considering the longer time frame of a whole phrase \cite{choi2017attention}. In an ASA context, this architecture learns to find pitch events at both the frame and phrase levels, and attention mechanisms help sequence models focus on frames with useful information. In audio classification, temporal-attention and attention-pooling are utilized to evaluate frames and reduce the significance level of less informative segments \cite{xu2017attentionaudio,lu2018temporal}. This is similar to assessment tasks where only a small number of frames show a big piece of event evidence, such as a wrong note or a change in pitch. Using embedding spaces in metric learning, the model identifies the difference between correct and incorrect examples. A common type uses triplet loss \cite{zhang2021ismir}. Triplet loss moves good examples closer together and bad examples farther apart in the learned space. This helps to provide an assessment without needing a strict scoring system for every singer and every piece. Using self-attention over longer contexts to engage with music error detection is a popular approach that has shown to be effective. Polytune allows transformers to work end-to-end, taking audio as input and generating annotated scores as output, and using synthetic data generation to scale training. The implemented mode achieved an average Error Detection \textit{F1} score of 64.1\% \cite{chou2025polytune}.

Transformers can interpret long-range dependencies of the input hence provide proper input to support error detection of musical contexts. ASA could become more effective by developing reference-independent evaluation methods. \newcite{gupta2018apsipa} developed an evaluation system that uses perceptual singing quality cues to rank songs without requiring a reference singer or melody. This method uses Best-Worst Scaling and twin neural networks to teach singing vocal ranking order in a preference-based system \cite{gupta2020twin}. The aim was to create methods that preserve system stability when playing different musical tracks and vocal artists because there are no established reference performances in actual learning environments.

A broader view of deep learning in singing assessment and related topics is summarized in a survey-style discussion of methods and tasks in singing voice analysis and assessment \cite{gupta2022taslp}. The overview section contains information about dataset restrictions and typical system breakdowns while emphasizing the requirement to connect model predictions with educational assessment results.

\subsection{Microtonal and Maqam-Based Studies}

A system concerned with microtonality must retain pitch data at an extremely granular level. In this context, research on Turkish \textit{makam} (Turkish word for maqam) offers useful frameworks that address pitch-based histogram analysis, tonic identification, and implementation of modal templates \cite{gedik2010hist,bozkurt2014review}. \newcite{gedik2010hist} details how analytical techniques that use pitch-frequency histograms can detect tonics and elucidate constructions of makam works . Their findings show that the recognition of the mode relies on the distribution of pitch, rather than on equality subdivisions of the octave into twelve bins. \newcite{bozkurt2014review} review computational methods to analyze Turkish makam, recognizing three challenges related to musical tuning, melodic progression, and the structure of rhythmic patterns. They identify three key needs: estimations of tonic reference, alignment of pitch distribution, and methods of evaluation, In addition to `practices of music theory' by its practitioners.

\newcite{yesiler2018makam} applied a Multi-layer Perceptron (MLP) classifier and features of extended pitch distribution to recognize Turkish makam. They achieved a mean accuracy of 0.756 within the framework of the evaluation setup they reported. The results showed that pitch features that are sensitive to microtones yielded better outcomes with respect to mode recognition than the traditional 12-tone quantization. Research on Maqam-related work shows that pitch-aware representations that use pitch distributions from continuous $f_0$ tracking produce better results. \newcite{yesiler2018makam} uses pitch-distribution features with a Multi-Layer Perceptron (MLP) classifier for makam recognition, showing that microtonal pitch representations outperform 12-tone quantized features.

Musicians need to identify particular intervals in Kurdish maqam music because these intervals specify the exact way to play their musical notes. The detailed contour tracing enables listeners to capture microtonal shifts by listening and learning with enough pitch detail 
\cite{gedik2010hist,bozkurt2014review}. This recognition process is continuous contour tracing defined by the pitch range intervals. Consequently, with respect to the recitation of Kurdish maqam, researchers analyze the microtonal documentation and pitch contour quality assessment for the first step of their analysis. Maqam studies require outcome symbolism for data design \cite{karaosmanoglu2012symbtr}.

\newcite{karaosmanoglu2012symbtr} supplies Turkish makam music as primary data with symbologies pertinent to Music Information Retrieval (MIR) tasks and metadata on makams, usuls (rhythmic cycles), and forms (structural compositions). Using the data, researchers can apply constrained musicological analyses and inquiries on the sequential arrangements of musical phrases and the structuring of songs. They can also validate inputs against the symbolic descriptions in order to retrieve and compare the audio results. \newcite{senturk2016thesis} analyzes Ottoman-Turkish makam music through audio and symbolic data analysis and applies tonic recognition and other computational functions.

The studies show that audio-based pitch-tracking employing modal theory allows researchers to build more sophisticated models as opposed to $f_0$ contour analysis \cite{gedik2010hist,bozkurt2014review}. Literature also shows international boundary discrepancies between boundary descriptors and the descriptors \cite{bozkurt2014review}. Recognition systems ascertain the pattern in question. Evaluation systems assess performance and offer commentary on the quality of performance. Classifying is limited for the learners. Feedback must explain the specific nature of the errors made in order to fulfill its purpose. Feedback must explain what it is that the learners did incorrectly, explanations that also include how this circumstantially relates to the rules or, in particular, the framework foundation of the mode \cite{gupta2022taslp}.

Most of the ASA research has been transferred from alignment and handcrafted features to statistical learning and deep learning. CNNs, recurrent models, attention methods, metric learning, and transformers broadened the scope of what systems audio learning from. Reference-independent evaluation and preference-based ranking works provide examples of bypassing just matching references \cite{gupta2018apsipa,gupta2020twin}. 64.1\% Error Detection F1 is reported in the Transformer-based error correction setting \cite{chou2025polytune}. The non-Western pitch systems in the Microtonal and maqam recognition research show that pitch distributions, tonic reference estimations, and microtonal-aware representations can be successfully modeled \cite{gedik2010hist,yesiler2018makam}. 

The singing of Maqam is one of the many singing styles in the Kurdish singing tradition that require microtonal intervals, movement of the pitch in fluid and differing manners, and the use of ornamental embellishments, which are all necessary for the performance to be considered correct and complete \cite{hashemi2021kurdishmaqams,shafiei2019ornamentation}. However, the ASA systems in the West consider these events as inaccuracies and failures within the performance. Scoring rubrics in ASA specific to Kurdish singing need to be created to align these systems with the values and practices seen in the Kurdish tradition, in addition to systems that are capable of tracking substantial changes in pitch and stability of the mode to be incorporated. A serious problem in the field is not having enough data. There are some annotated Kurdish singing datasets, but they are hard to find and expensive to make because they need expert annotation. Because of this, models that are trained on one regional singing style often do not work with other styles, even though all of them are valid. Another gap relates to feedback design. Many systems output numerical scores or pitch traces. In the pedagogy of singing in Kurdish culture, feedback is tailored to specific descriptive comments that align with particular modal behaviors and the use of particular ornaments \cite{gupta2022taslp}. The gap of translating model output into specific feedback guidance remains nuanced when considering data-driven pedagogy in the Kurdish context.
As a result, the Kurdish-centered ASA for maqam should focus on microtonal pitch tracking, maqam-aware structure, and Kurdish labels.

\section{Method}
\begin{figure}[ht!]
\centering
\input{figures/research_design_tikz.tex}
\caption{Overview of the research design we decided to use in this study.}
\label{fig:research_design}
\end{figure}
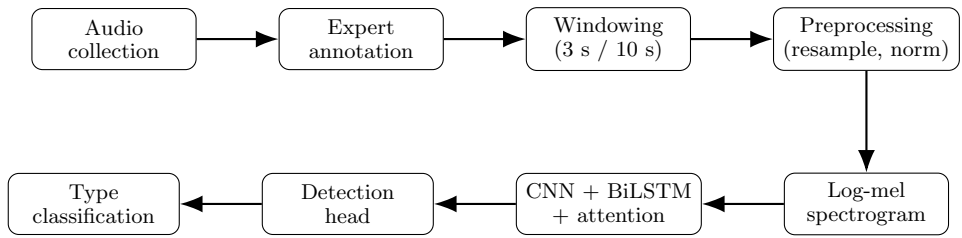

Figure~\ref{fig:research_design} shows an overview of the research method. We record Bayati-kurd maqam sang by Kurdish singers and prepare it for annotation by experts. The annotators identify the beginning and ending timestamps of fine pitch errors, rhythm errors, and modal drift errors. The annotated data is used to prepare a dataset to train a model based on a combination of  CNN and BiLSTM with attention. 

The annotated data are divided into small segments (windows), each associated with labels indicating what errors are present and what kind of errors they are. The model learns patterns directly from the data. Each window is converted into a log-mel spectrogram, a representation that captures changes in time-varying frequencies and is widely used in audio learning and music signal processing \cite{muller2015fundamentals}.

\vspace{-12pt}

\begin{figure}[H]
\centering
\setlength{\fboxrule}{0.6pt}
\setlength{\fboxsep}{0pt}
\begin{subfigure}[t]{0.32\textwidth}
\centering
\fbox{\includegraphics[width=\dimexpr\linewidth-1.2pt\relax]{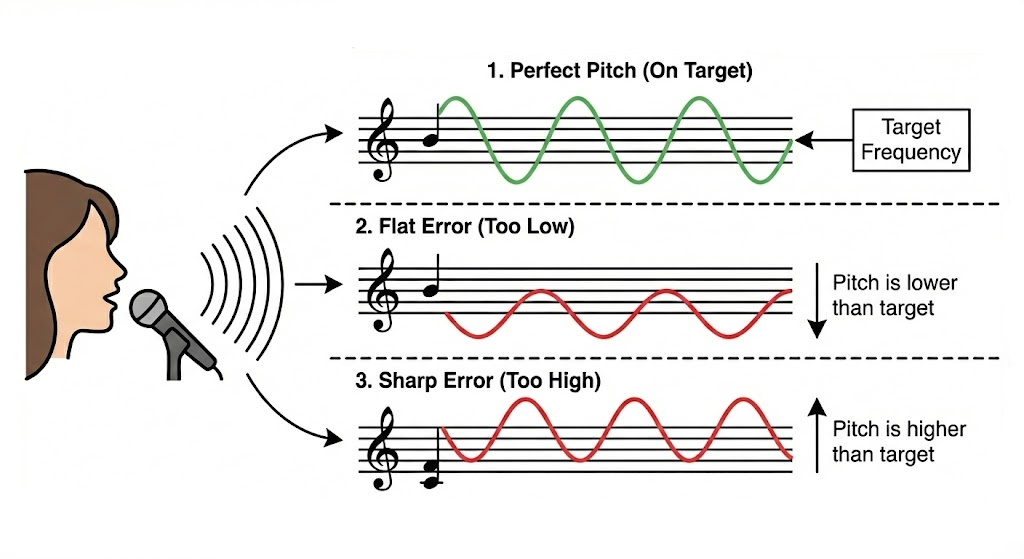}}
\caption{Fine pitch error.}
\label{fig:fine_pitch_error}
\end{subfigure}\hfill
\begin{subfigure}[t]{0.32\textwidth}
\centering
\fbox{\includegraphics[width=\dimexpr\linewidth-1.2pt\relax]{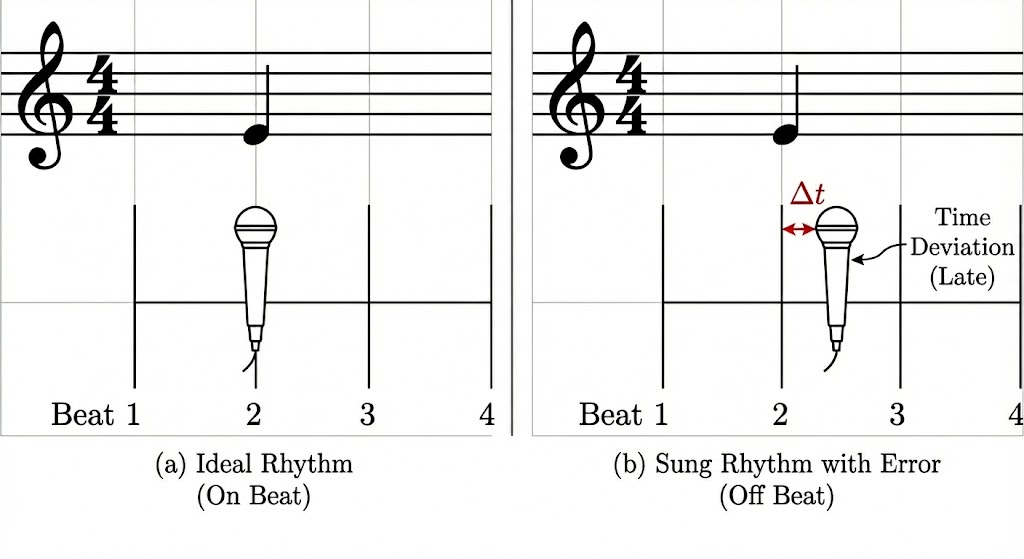}}
\caption{Rhythm error.}
\label{fig:rhythm_error}
\end{subfigure}\hfill
\begin{subfigure}[t]{0.32\textwidth}
\centering
\fbox{\includegraphics[width=\dimexpr\linewidth-1.2pt\relax]{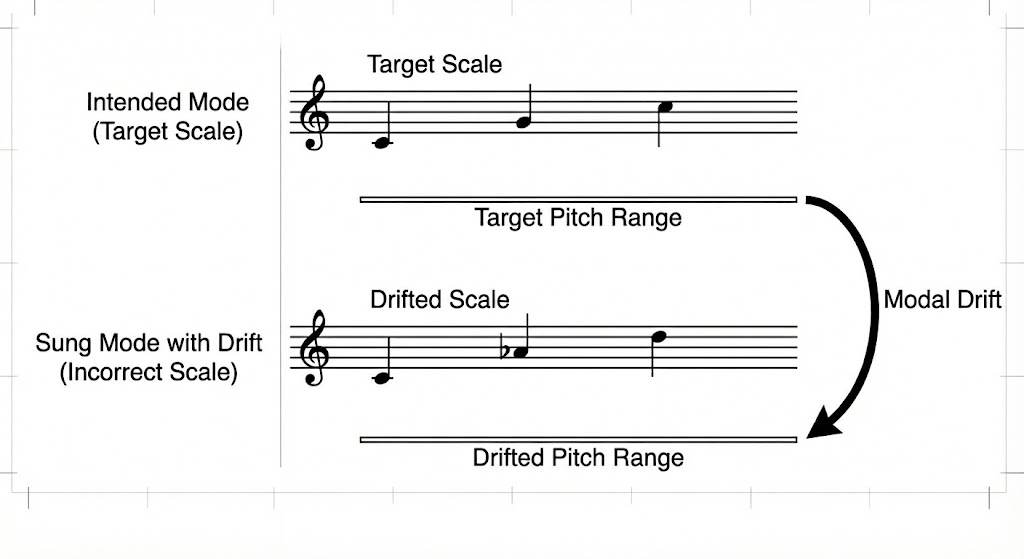}}
\caption{Modal drift.}
\label{fig:modal_drift_error}
\end{subfigure}
\caption{The three error types: (a) fine pitch; (b) rhythm; (c) modal drift.}
\vspace{-4pt}
\end{figure}

The songs, from both female and male singers, are recorded at 22,050 Hz (WAV, mono). In this research, we ask the annotators only to focus on three error types: fine pitch, rhythm, and modal drift. Figures~\ref{fig:fine_pitch_error}, \ref{fig:rhythm_error}, and \ref{fig:modal_drift_error} illustrate these types.

Audio is converted to log-mel spectrograms (1024 FFT, 512 hop, 128 mel bins) \cite{hershey2017cnn,pons2016cnn} and normalised \cite{ioffe2015batchnorm}. We use 10 s windows (1 s hop) and 3 s windows (0.5 s hop) from training songs only; labels follow a centre-overlap rule (20\% centre, 30\% overlap).

The model has two main components: a CNN that extracts local patterns from the spectrogram (following work by \newcite{bittner2017deep} that applies CNNs to melody extraction and pitch-related tasks), and a BiLSTM to model temporal structure and longer musical context. We also add an attention mechanism that focuses on the most important time steps \cite{choi2017attention}. We design the model to have two output heads. One determines whether an error exists in the window, and the other determines the type of error \cite{caruana1997multitask}. The outputs are evaluated against the ground truth labels during training. We use weighted sampling and a loss function that assigns more weight to difficult samples to help reduce the effect of class imbalance (based on the work by \newcite{lin2017focalloss}).

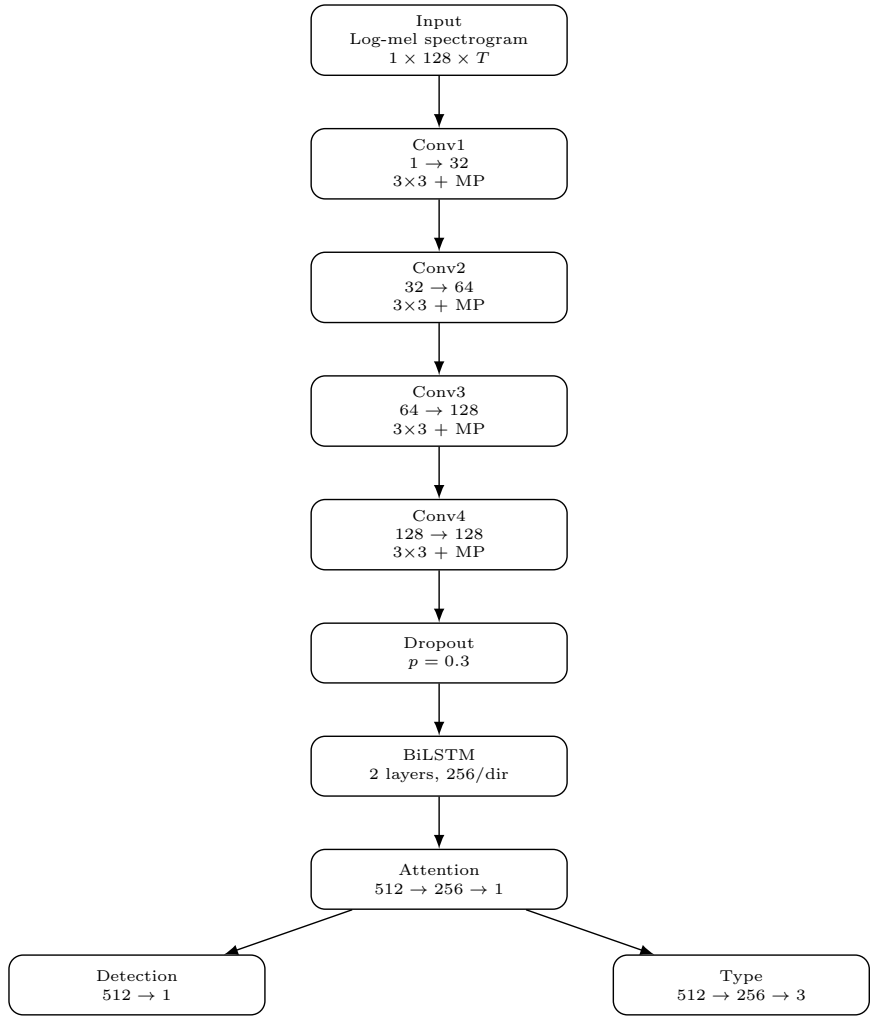
\begin{figure}[H]
\centering
\resizebox{0.72\textwidth}{!}{%
  \input{figures/model_architecture_tikz.tex}%
}
\caption{Model architecture with CNN feature extractor, bidirectional LSTM encoder, attention pooling, and two-task output heads.}
\label{fig:model_architecture}
\vspace{-4pt}
\end{figure}

The model is a CNN-BiLSTM with attention and two heads: detection (sigmoid) and type classification (softmax over three classes). Figure~\ref{fig:model_architecture} and Table~\ref{tab:model_architecture} summarize the architecture.

\begin{table}[ht!]
\centering
\caption{Summary of the model architecture.}
\label{tab:model_architecture}
\renewcommand{\arraystretch}{1.2}
\begin{tabularx}{0.88\linewidth}{|l|X|l|}
\hline
\rowcolor{lightgray}
\cellcolor{lightgray}\textbf{Component} & \cellcolor{lightgray}\textbf{Details} & \cellcolor{lightgray}\textbf{Output size} \\
\hline
Input & Log-mel spectrogram & $1 \times 128 \times T$ \\
\hline
CNN Blocks 1--4 & 3$\times$3 Conv, BN, ReLU, MaxPool 2$\times$2 & $128 \times 8 \times T/16$ \\
\hline
BiLSTM & 2 layers, 256 per direction, dropout 0.3 & Sequence of 512 dims \\
\hline
Attention & 512$\rightarrow$256 (Tanh), dropout 0.2, 256$\rightarrow$1, softmax & 512 dims \\
\hline
Detection head & Dropout 0.3, Linear 512$\rightarrow$1, sigmoid & 1 (probability) \\
\hline
Type head & Linear 512$\rightarrow$256, ReLU, dropout 0.3, Linear 256$\rightarrow$3, softmax & 3 classes \\
\hline
\end{tabularx}
\end{table}

We train the model with AdamW \cite{loshchilov2019decoupled} ($3\times10^{-4}$, weight decay $2\times10^{-4}$), ReduceLROnPlateau, weighted~binary cross-entropy for detection, focal loss ($\gamma=2.0$) for type (based on what \newcite{lin2017focalloss} proposed), class and $\alpha$-weights ($\alpha_{\text{modal drift}}=5.0$), batch size~16, early stopping (patience~10), and augmentation (time stretch, pitch shift, noise, gain, SpecAugment) (according to \newcite{park2019specaugment}) plus~hard negative mining (following \newcite{shrivastava2016training}).

\section{Results and Discussion}

This section presents the experimental setup, the results of data collection, data preparation, model training, evaluation, and analysis of findings.

\subsection{Experimental Setup}

The initial setup aimed to address the challenges of detecting and categorizing vocal errors present in Kurdish microtonal singing by accounting for class imbalance and maximizing the model's efficiency. Over time, the experimental setup changed in an iterative fashion, allowing new challenges to be identified and resolved in each stage.

\subsubsection{Annotation Tool Development}

\begin{figure}[H]
\centering
\setlength{\fboxrule}{0.6pt}
\setlength{\fboxsep}{0pt}
\fbox{\includegraphics[width=0.85\textwidth]{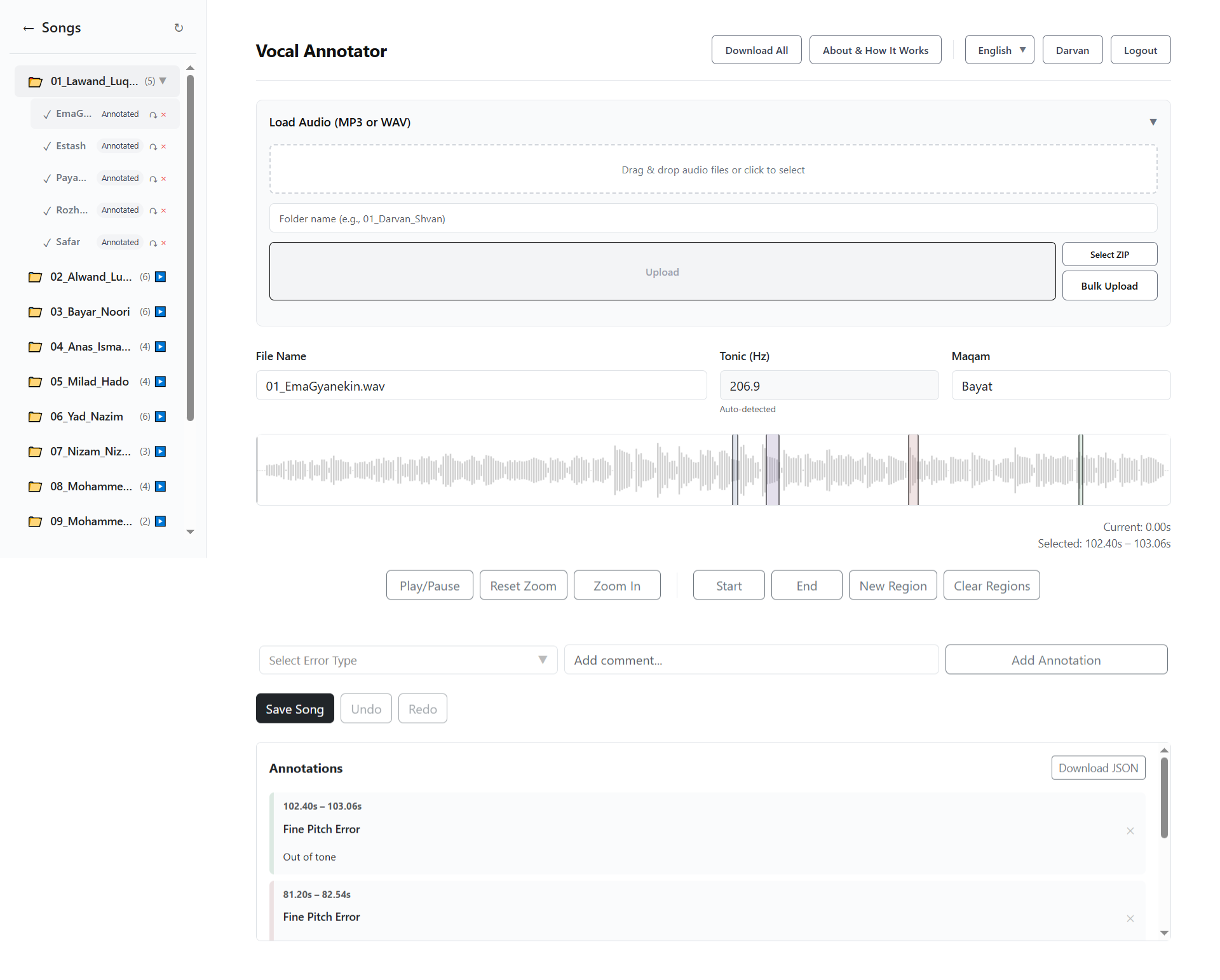}}
\caption{The Vocal Annotator interface used for creating training annotations.}
\label{fig:annotator_home}
\end{figure}

To facilitate the annotation process, we developed a custom web-based tool, \textit{Vocal Annotator}. It offers interfaces in Kurdish Sorani, Kurdish Kurmanji, English, Arabic, Persian, and Turkish. The annotators, from different backgrounds, can use the application to annotate the data. It provides automatic tonic detection, waveform display with region selection at millisecond resolution for marking error spans, project organization by singer, and export of annotations in JSON for direct use in the model training pipeline. Its user-friendly design makes it easy for musicians and annotators without any computational background to do their annotation task efficiently. Figure~\ref{fig:annotator_home} shows the Vocal Annotator interface.

Annotators were given a clear guideline: use headphones, mark only audible errors (not stylistic choices), and keep regions tight around each error. The annotation process used a three-level validation scheme to ensure the label quality. Level~1: The primary researcher produced initial annotations. Level~2: A music student performed an independent pass. Level~3: A Kurdish maqam expert reviewed all annotations for correctness and consistency with Bayati-kurd practice. This multi-level validation reduced individual bias and improved the reliability of the training labels. Each annotation file was exported as JSON for the training dataset; an example is given in Figure~\ref{fig:json_example}. 

\begin{figure}[H]
\centering
\begin{minipage}{0.75\linewidth}
\begin{verbatim}
[
  {
    "start": 102.404,
    "end": 103.064,
    "type": "fine_pitch_error",
    "detail": "Out of tone"
  },
  {
    "start": 81.196,
    "end": 82.536,
    "type": "fine_pitch_error",
    "detail": "Out of tone"
  }
]
\end{verbatim}
\end{minipage}
\caption{Example JSON annotation exported from the Vocal Annotator.}
\label{fig:json_example}
\end{figure}

\subsection{Collected Data}

\begin{figure}[ht!]
\centering
\input{figures/error_distribution_pie.tex}
\caption{Error type distribution in the collected dataset (221 annotations).}
\label{fig:error_pie}
\end{figure}
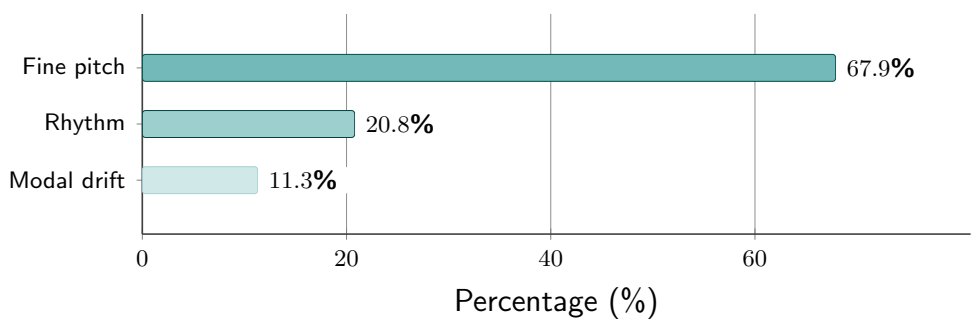

Recordings were collected with informed consent from all participants. The data collection stage produced a focused set of Kurdish microtonal singing recordings in the Bayati-kurd maqam. Each file includes a vocal performance and a set of manual labels that mark vocal mistakes. There are 50 songs in total, recorded by 13 singers, three female and ten male. The number of songs per singer is uneven, which is normal for this kind of material. One singer contributed six songs, five singers contributed five or six songs, four singers contributed four songs each, two singers contributed three songs each, and one singer recorded a single piece. The total duration across the 50 songs is approximately two and a half hours. The files are stored in WAV format with a sample rate of 22,050 Hz. Most of the recordings run between one and three minutes. The annotation yielded 221 vocal errors across all songs, resulting in an average of about 4.4 errors per song. Each label is stored as a JSON entry containing the error type, start time, end time, and an optional short note. The three error types are heavily unbalanced. Fine pitch errors are the largest group with 150 examples, which is 67.9 percent of all errors. Rhythm errors follow with 46 examples, 20.8 percent. Modal drift is rare, having only 25 examples, presenting 11.3\% of the total. This is roughly a 6.0 to 1.8 to 1.0 ratio. Modal drift appears in only 8 songs. This small spread makes it hard for any model to learn it. Table~\ref{tab:collected_data} and Table~\ref{tab:error_distribution} provide the dataset overview and error-type counts. Figure~\ref{fig:error_pie} illustrates the error-type distribution.

The annotation took about sixty hours, reflecting the difficulty of identifying microtonal pitch shifts, rhythmic changes, and modal behavior in Bayati-kurd. 

\begin{table}[ht!]
\centering
\caption{Dataset collection statistics.}
\label{tab:collected_data}
\renewcommand{\arraystretch}{1.2}
\begin{tabularx}{0.88\linewidth}{|l|>{\centering\arraybackslash}X|}
\hline
\rowcolor{lightgray}
\cellcolor{lightgray}\textbf{Statistic} & \cellcolor{lightgray}\textbf{Value} \\
\hline
Total songs & 50 \\
Total singers & 13 \\
Total annotations & 221 \\
Average errors per song & 4.4 \\
Songs with modal drift & 8 \\
\hline
\end{tabularx}
\end{table}

\begin{table}[ht!]
\centering
\caption{Error type distribution in the collected dataset.}
\label{tab:error_distribution}
\renewcommand{\arraystretch}{1.2}
\begin{tabularx}{0.88\linewidth}{|l|>{\centering\arraybackslash}X|>{\centering\arraybackslash}X|}
\hline
\rowcolor{lightgray}
\cellcolor{lightgray}\textbf{Error type} & \cellcolor{lightgray}\textbf{Count} & \cellcolor{lightgray}\textbf{Percentage} \\
\hline
Fine pitch error & 150 & 67.9\% \\
Rhythm error & 46 & 20.8\% \\
Modal drift error & 25 & 11.3\% \\
\hline
Total & 221 & 100\% \\
\hline
\end{tabularx}
\end{table}

\subsection{Dataset Preparation}

The audio and labels went through preprocessing to prepare them for deep learning. This stage included audio formatting, feature extraction, window creation, and splitting into training, validation, and test sets. All audio files were loaded at 22,050 Hz, converted to mono, and peak-normalized. We turned each file into a log-mel spectrogram with 128 mel bands, an FFT size of 1024, and a hop length of 512 samples. The spectrogram values were converted to the decibel scale and standardized using the mean and standard deviation. The model uses sliding windows: long windows of 10 seconds with a hop of one second give a broad context; short windows of three seconds with a hop of 0.5 seconds were produced only from training songs, adding more detailed examples, which together produced 15,199 windows. Labels follow a center-based rule. A window is marked as containing an error when the annotation overlaps at least 30 percent with the center 20 percent of the window. Windows with no overlap are marked as clean. The split is done by song, not by window, to prevent leakage. The result is 13,363 windows for training, 718 for validation, and 1,118 for testing. Class ratios remain similar in each split. In the training subset, 796 windows contain errors, and 12,567 are clean. Fine pitch errors appear in 537 windows, rhythm errors in 212 windows, and modal drift in only 47 windows. Table~\ref{tab:dataset_split} and Table~\ref{tab:training_errors} summarize the split and error distribution in the training set.

\begin{table}[H]
\centering
\caption{Dataset split statistics.}
\label{tab:dataset_split}
\renewcommand{\arraystretch}{1.2}
\begin{tabularx}{0.88\linewidth}{|l|>{\centering\arraybackslash}X|>{\centering\arraybackslash}X|}
\hline
\rowcolor{lightgray}
\cellcolor{lightgray}\textbf{Split} & \cellcolor{lightgray}\textbf{Windows} & \cellcolor{lightgray}\textbf{Percentage} \\
\hline
Training & 13,363 & 88.0\% \\
Validation & 718 & 4.7\% \\
Test & 1,118 & 7.4\% \\
\hline
Total & 15,199 & 100\% \\
\hline
\end{tabularx}
\end{table}

\begin{table}[H]
\centering
\caption{Error distribution in training set.}
\label{tab:training_errors}
\renewcommand{\arraystretch}{1.2}
\begin{tabularx}{0.88\linewidth}{|l|>{\centering\arraybackslash}X|>{\centering\arraybackslash}X|}
\hline
\rowcolor{lightgray}
\cellcolor{lightgray}\textbf{Error type} & \cellcolor{lightgray}\textbf{Count} & \cellcolor{lightgray}\textbf{Percentage} \\
\hline
Fine pitch error & 537 & 67.5\% \\
Rhythm error & 212 & 26.6\% \\
Modal drift error & 47 & 5.9\% \\
\hline
Total errors & 796 & -- \\
Clean windows & 12,567 & 94.0\% \\
\hline
\end{tabularx}
\end{table}

\subsection{Training Results}

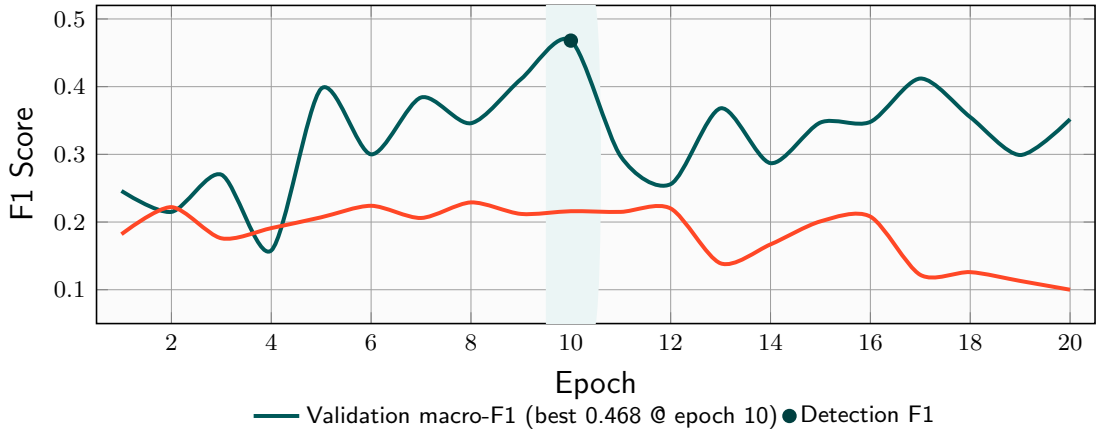
\begin{figure}[H]
\centering
\input{figures/training_curves.tex}
\caption{Validation macro-F1 and detection F1 over 20 epochs. Best validation macro-F1 (0.468) at epoch 10.}
\label{fig:training_curves}
\end{figure}

The training was done on a single GPU (GTX 1050 initially, then RTX 5090). We fixed the random seed (42) for reproducibility. The system saved three kinds of checkpoints: the best model when validation macro-F1 improved, periodic checkpoints every five epochs, and an emergency checkpoint if training was stopped manually. Training took place over 20 epochs. Early stopping chose epoch 10 as the best point. The run took 1 to 2 hours on an NVIDIA RTX 5090 at over 1,600 samples per second. Loss of detection started at about 2.0 and dropped to about 1.7--1.8 by epoch 5. The loss of classification started at 0.8 and went down to about 0.2--0.3 by epoch 10. Detection accuracy on training data went from about 20--30\% to about 40--50\% by epoch 10. Type prediction accuracy on windows with errors went from 30--40\% to 60--70\%. Type prediction was more accurate than detection throughout. The best validation macro-F1 was 0.468 and validation detection F1 was 0.216 at epoch 10. Fine pitch errors improved quickly; rhythm improved more slowly; modal drift stayed low with high variance, matching class sizes (537, 212, 47). One strong pattern was the gap between detection and classification: once a window is flagged, fine pitch labels were correct 89.5\% of the time and rhythm 76.0\%, confirming that finding error windows is harder than classifying their type. Table~\ref{tab:evaluation_results} and Table~\ref{tab:per_class_results} summarize the evaluation metrics; Figure~\ref{fig:training_curves} shows validation macro-F1 and detection F1 over epochs; the best checkpoint was at epoch 10.

\begin{table}[H]
\centering
\caption{Comprehensive evaluation results on all 50 songs.}
\label{tab:evaluation_results}
\renewcommand{\arraystretch}{1.2}
\begin{tabularx}{0.88\linewidth}{|l|>{\centering\arraybackslash}X|}
\hline
\rowcolor{lightgray}
\cellcolor{lightgray}\textbf{Metric} & \cellcolor{lightgray}\textbf{Value} \\
\hline
Detection threshold & 0.750 \\
Detection recall & 39.4\% \\
Detection precision & 25.8\% \\
Detection F1 & 0.311 \\
Type macro-F1 & 0.387 \\
\hline
\end{tabularx}
\end{table}

\begin{table}[H]
\centering
\caption{Per-class classification performance on all 50 songs.}
\label{tab:per_class_results}
\renewcommand{\arraystretch}{1.2}
\begin{tabularx}{0.88\linewidth}{|l|>{\centering\arraybackslash}X|>{\centering\arraybackslash}X|>{\centering\arraybackslash}X|>{\centering\arraybackslash}X|}
\hline
\rowcolor{lightgray}
\cellcolor{lightgray}\textbf{Error type} & \cellcolor{lightgray}\textbf{Detected} & \cellcolor{lightgray}\textbf{Precision} & \cellcolor{lightgray}\textbf{Recall} & \cellcolor{lightgray}\textbf{F1 score} \\
\hline
Fine pitch error & 57 & 89.5\% & 34.0\% & 0.492 \\
Rhythm error & 25 & 76.0\% & 41.3\% & 0.536 \\
Modal drift error & 5 & 40.0\% & 8.0\% & 0.133 \\
\hline
\end{tabularx}
\end{table}

\subsection{Evaluation Results}

To avoid bias from a small test split, the final model was evaluated on all 50 songs, covering all 221 ground truth errors. The detection threshold was set to 0.750. The detection head found 87 of 221 true errors (recall 39.4\%), produced 337 predictions in total, with 250 false positives (precision 25.8\%), giving detection F1 0.311. Type prediction was evaluated only on the 87 matched detections. Macro-F1 across the three classes was 0.387. For fine pitch: precision 89.5\%, F1 0.492. Rhythm: F1 0.536, precision 76.0\%, recall 41.3\%. Modal drift: F1 0.133, only 5 detections out of 25 true cases (2 correctly typed). Table~\ref{tab:evaluation_results} summarizes these overall metrics; Table~\ref{tab:per_class_results} gives per-class performance. A threshold search between 0.30 and 0.90 showed that 0.750 gave the best F1 on validation; lower thresholds gave high recall, but many false alarms, and higher thresholds missed many errors. The detection head has a recall of 39.4\% (87/221) and a precision of 25.8\% (87/337). The F1 of 0.311 reflects how hard it is to find errors when clean segments dominate. The 250 false positives show sensitivity to spectrogram changes that may not be real errors; the 134 false negatives show missed errors. The classification head puts each detected window into one of three categories. Fine pitch has the highest precision when detected (89.5\%); rhythm has the best overall F1 (0.536) with balanced precision and recall; modal drift is hardest (8.0\% recall, 0.133 F1, 40\% type accuracy when found). Figure~\ref{fig:confusion_viz} shows the confusion matrix for the 87 matched detections.

\begin{figure}[H]
\centering
\input{figures/confusion_matrix_viz.tex}
\caption{Confusion matrix for error-type classification (87 matched detections). Diagonal cells (shaded) are correct predictions.}
\label{fig:confusion_viz}
\end{figure}

\subsection{Analysis and Discussion}

Fine pitch: 51 correct, 4 as rhythm, 2 as modal. Rhythm: 19 correct, 6 as fine pitch. Modal drift: 2 correct out of 5 detected, 2 as fine pitch, 1 as rhythm. The main confusions are rhythm$\rightarrow$fine pitch (6) and fine pitch$\rightarrow$rhythm (4). Rhythmic instability can resemble pitch instability in the spectrogram; modal drift is so rare that the model often misclassifies it as fine pitch or rhythm.

The results show that the model is good at finding and sorting out small pitch and rhythm mistakes. The fine pitch error class has high type classification accuracy (89.5\%) when detected. Rhythm errors have the best overall F1 score (0.536). The modal drift error class is the hardest, with only 8.0\% recall and 0.133 F1, directly related to how rare modal drift examples are (25 across 8 songs). Detection performance is conservative due to the optimal threshold of 0.750. The confusion between fine pitch and rhythm errors suggests that these error types share similar spectro-temporal characteristics. Overall, the two-headed model can find and categorize the most common singing mistakes in the dataset; more data, especially for modal drift, would likely improve performance.

Type prediction works better than detection once a window is flagged. When the model finds an error window, it usually predicts the type correctly for fine pitch and rhythm. Their precision scores (89.5\% and 76.0\%) show that the shared CNN-BiLSTM backbone produces features that help with both tasks. Modal drift is so rare that the model does not have enough examples to learn it well, leading to misclassifications as the more common error types. Many false positives appear in expressive movements, fast transitions, quiet moments with noise, or pitch gestures that resemble errors in the spectrogram. Missed errors come from very small deviations, events near window borders (dropped by the centre-based rule), and modal drift (20 of 25 drift cases missed). Class-weighted loss, focal loss, weighted sampling, and targeted augmentation help but cannot fully compensate for the lack of modal drift data. The model can give feedback on mistakes in pitch and rhythm when singing Bayati-kurd maqam, which is useful in practice. It can point out parts that need attention and give hints about type. We still need human review, especially for drift errors and for richer feedback. The evaluation justifies future work: more modal drift examples, rhythm- and pitch-aware features to reduce confusion, and integration into an interactive tool with user studies.

\section{Conclusion and Future Work}

This research demonstrates that a two-head AI pipeline can deliver practical feedback for Kurdish microtonal singing in the Bayati-Kurd maqam. A curated corpus of 50 WAV recordings (22{,}050 Hz) from 13 vocalists produced 15{,}199 windows for training and 221 annotated error spans (25 modal drift), providing the first structured dataset for this task. Trained for 20 epochs with early stopping at epoch 10, the model reached validation macro-F1 of 0.468 and detection F1 of 0.216. On the full-song evaluation with an operating threshold of 0.750, detection recall was 39.4\% and precision 25.8\% (F1 = 0.311). Type prediction outperforms detection once a window is flagged: macro-F1 reached 0.387, with per-class F1 of 0.492 for fine pitch, 0.536 for rhythm, and 0.133 for modal drift. The precision for the frequent classes inside detected windows was high (fine pitch 89.5\%, rhythm 76\%), while modal drift remained recall-poor (8.0\%). The system is already useful in a human-in-the-loop coaching workflow for surfacing pitch and rhythm issues in Kurdish microtonal singing, and it establishes quantitative baselines for future improvements.

In the future, we are interested in expanding modal drift coverage to at least 100--150 examples across more singers and sessions; diversifying recording conditions and adding other maqams; including rhythm-aware and pitch-aware features to minimize confusion between fine pitch and rhythm; trying contrastive or metric losses; keeping focal loss and balanced sampling with targeted augmentation for modal drift; porting the pipeline to more maqams and to related vocal-analysis tasks; and finally, exploring lightweight or streaming variants for near-real-time feedback.

\section*{Acknowledgment}

We would like to express our sincere gratitude to Mr. Lawand Luqman for the manual annotation of vocal errors in the Bayati-kurd recordings, and to Mr. Hoshyar Karim for checking the annotations. We also thank Mr. Wirya Ahmed for providing additional musical instructions. We are deeply grateful to the volunteer singers who contributed their recordings to this project: Mr. Matin Shvan, Mr. Lawand Luqman, Mr. Alwand Luqman, Mr. Bayar Noori, Mr. Anas Ismahel, Mr. Milad Hado, Mr. Yad Nazim, Mr. Nizam Nizamadeen, Mr. Mohammed Nabil, Mr. Mohammed Dilshad, Ms. Shayan Shukur, Ms. Rayan Rahman, Ms. Ava Shawkat, Ms. Khanda Abdullah, Ms. Shagul Salar. We also appreciate the support of our colleagues and the Revge team for the friendly environment and steady teamwork. 

\section*{Ethical Consideration}

The participants provided their consent to use and publicize the recordings. The publicly released dataset associated with this study has been processed to remove any sensitive personal information that could be traced back to any particular individual to minimize potential risks to individuals whose tweets were included.

\section*{Author Contributions}

Conceptualization, Hossein Hassani (H.H.) and Darvan Shvan Khairaldeen (D.S.K.); methodology, D.S.K. and H.H; software, D.S.K.; validation, D.S.K. and H.H.; formal analysis, D.S.K.; investigation, D.S.K.; resources, D.S.K. and H.H.; data curation, D.S.K.; preparing first draft, D.S.K.; revising the draft and preparing final manuscript, D.S.K.; visualization, D.S.K.; supervision, D.S.K.; project administration, H.H.

\section*{Data Availability}

The data will be publicized on `data.krd` and `KurdishBLARK`, but it may take some time before it becomes accessible. 

\bibliographystyle{lrec}
\bibliography{main}

\end{document}

%% file: figures/research_design_tikz.tex
\begin{tikzpicture}[
    scale=0.9,
    transform shape,
    node distance=1.2cm,
    every node/.style={font=\small},
    box/.style={
        rectangle,
        draw,
        rounded corners,
        align=center,
        minimum width=2.4cm,
        minimum height=0.9cm
    },
    arrow/.style={
        -{Latex[length=3mm]},
        thick
    }
]

\node[box] (record) {Audio\\collection};
\node[box, right=of record] (annot) {Expert\\annotation};
\node[box, right=of annot] (window) {Windowing\\(3 s / 10 s)};
\node[box, right=of window] (prep) {Preprocessing\\(resample, norm)};

\node[box, below=1.5cm of prep] (feat) {Log-mel\\spectrogram};
\node[box, left=of feat, minimum width=2.7cm] (model) {CNN + BiLSTM\\+ attention};
\node[box, left=of model, minimum width=2.5cm] (detect) {Detection\\head};
\node[box, left=of detect, minimum width=2.5cm] (type) {Type\\classification};

\draw[arrow] (record) -- (annot);
\draw[arrow] (annot) -- (window);
\draw[arrow] (window) -- (prep);

\draw[arrow] (prep) -- (feat);

\draw[arrow] (feat) -- (model);
\draw[arrow] (model) -- (detect);
\draw[arrow] (detect) -- (type);

\end{tikzpicture}

%% file: figures/model_architecture_tikz.tex
\begin{tikzpicture}[
    node distance=7mm,
    >=latex,
    scale=0.75,
    every node/.style={transform shape},
    block/.style={
        draw,
        rounded corners,
        align=center,
        minimum width=34mm,
        minimum height=8mm,
        font=\tiny
    }
]

\node[block] (input)
{Input\\Log-mel spectrogram\\$1 \times 128 \times T$};

\node[block, below=of input] (cnn1)
{Conv1\\$1 \rightarrow 32$\\3×3 + MP};

\node[block, below=of cnn1] (cnn2)
{Conv2\\$32 \rightarrow 64$\\3×3 + MP};

\node[block, below=of cnn2] (cnn3)
{Conv3\\$64 \rightarrow 128$\\3×3 + MP};

\node[block, below=of cnn3] (cnn4)
{Conv4\\$128 \rightarrow 128$\\3×3 + MP};

\node[block, below=of cnn4] (drop1)
{Dropout\\$p=0.3$};

\node[block, below=of drop1] (lstm)
{BiLSTM\\2 layers, 256/dir};

\node[block, below=of lstm] (attn)
{Attention\\$512 \rightarrow 256 \rightarrow 1$};

\node[block, below left=6mm and 6mm of attn] (det)
{Detection\\$512 \rightarrow 1$};

\node[block, below right=6mm and 6mm of attn] (cls)
{Type\\$512 \rightarrow 256 \rightarrow 3$};

\draw[->] (input) -- (cnn1);
\draw[->] (cnn1) -- (cnn2);
\draw[->] (cnn2) -- (cnn3);
\draw[->] (cnn3) -- (cnn4);
\draw[->] (cnn4) -- (drop1);
\draw[->] (drop1) -- (lstm);
\draw[->] (lstm) -- (attn);
\draw[->] (attn) -- (det);
\draw[->] (attn) -- (cls);

\end{tikzpicture}

%% file: figures/error_distribution_pie.tex
\begin{tikzpicture}
\begin{axis}[
    width=0.78\textwidth,
    height=4.5cm,
    xlabel={Percentage (\%)},
    xlabel style={font=\sffamily},
    ylabel style={font=\sffamily},
    tick label style={font=\sffamily\small},
    xmin=0, xmax=78,
    xtick={0,20,40,60},
    ytick={0,1,2},
    yticklabels={Modal drift, Rhythm, Fine pitch},
    yticklabel style={font=\sffamily\small},
    xbar,
    bar width=0.48,
    bar shift=0pt,
    axis lines=left,
    axis line style={-, line width=0.6pt, draw=black!75},
    x axis line style={-},
    y axis line style={-},
    axis background/.style={fill=white},
    xmajorgrids=true,
    ymajorgrids=false,
    grid style={line width=0.25pt, draw=gray!88},
    nodes near coords={\pgfmathprintnumber[fixed,precision=1]{\pgfplotspointmeta}\%},
    nodes near coords align={horizontal},
    every node near coord/.append style={
      font=\sffamily\small\bfseries,
      inner xsep=4pt,
      inner ysep=1.5pt,
      anchor=west,
      fill=white,
      draw=none,
    },
    enlarge y limits=0.48,
    enlarge x limits={value=0.04,upper},
]
\addplot[fill=teal!55, draw=teal!60!black, line width=0.35pt, rounded corners=1.2pt]
  coordinates {(67.9,2)};
\addplot[fill=teal!38, draw=teal!52!black, line width=0.35pt, rounded corners=1.2pt]
  coordinates {(20.8,1)};
\addplot[fill=teal!18, draw=teal!35, line width=0.35pt, rounded corners=1.2pt]
  coordinates {(11.3,0)};
\end{axis}
\end{tikzpicture}

%% file: figures/training_curves.tex
\begin{tikzpicture}
\begin{axis}[
    width=0.92\textwidth,
    height=5.8cm,
    xlabel={Epoch},
    ylabel={F1 Score},
    xlabel style={font=\sffamily},
    ylabel style={font=\sffamily},
    tick label style={font=\sffamily\small},
    legend style={
      at={(0.5,-0.22)},
      anchor=north,
      legend columns=2,
      font=\sffamily\footnotesize,
      draw=none,
      fill=none,
      cells={anchor=west}
    },
    grid=both,
    grid style={line width=0.2pt, draw=gray!85},
    major grid style={line width=0.35pt, draw=gray!75},
    xmin=0.5, xmax=20.5,
    ymin=0.05, ymax=0.52,
    xtick={0,2,...,20},
    ytick={0.1,0.2,...,0.5},
    axis background/.style={fill=gray!3},
    axis line style={line width=0.6pt, draw=black},
    every axis plot/.append style={line width=1.4pt, smooth},
]
\addplot[draw=none, fill=teal!8, forget plot] coordinates {(9.5,0.05) (10.5,0.05) (10.5,0.52) (9.5,0.52)} \closedcycle;
\addplot[color=teal!70!black, mark=none, smooth, line width=1.5pt]
  coordinates {
    (1,0.246) (2,0.215) (3,0.270) (4,0.158) (5,0.397)
    (6,0.300) (7,0.384) (8,0.346) (9,0.411) (10,0.468)
    (11,0.297) (12,0.256) (13,0.368) (14,0.287) (15,0.347)
    (16,0.348) (17,0.412) (18,0.355) (19,0.299) (20,0.352)
  };
\addplot[color=teal!50!black, mark=*, only marks, mark size=2pt]
  coordinates {(10,0.468)};
\addlegendentry{Validation macro-F1 (best 0.468 @ epoch 10)}
\addplot[color=red!70!orange!85, mark=none, smooth, line width=1.5pt]
  coordinates {
    (1,0.182) (2,0.222) (3,0.176) (4,0.191) (5,0.207)
    (6,0.224) (7,0.206) (8,0.229) (9,0.212) (10,0.216)
    (11,0.215) (12,0.220) (13,0.139) (14,0.167) (15,0.201)
    (16,0.208) (17,0.122) (18,0.126) (19,0.113) (20,0.100)
  };
\addlegendentry{Detection F1}
\end{axis}
\end{tikzpicture}

%% file: figures/confusion_matrix_viz.tex
\begin{tikzpicture}[
  font=\sffamily,
  cell/.style={
    rectangle,
    minimum width=1.6cm,
    minimum height=1.05cm,
    inner sep=0pt,
    outer sep=0pt,
    align=center,
  },
  header/.style={cell, font=\sffamily\bfseries\small, text=white, fill=teal!50!black},
]
  \matrix (M) [matrix of nodes, row sep=3pt, column sep=3pt, nodes={cell, anchor=center}]
  {
    |[header]| & |[header]| Fine & |[header]| Rhythm & |[header]| Modal \\
    |[header]| Fine   & |[fill=teal!55]| \textbf{51} & |[fill=teal!8]| 4 & |[fill=teal!6]| 2 \\
    |[header]| Rhythm & |[fill=teal!12]| 6 & |[fill=teal!38]| \textbf{19} & |[fill=teal!5]| 0 \\
    |[header]| Modal  & |[fill=teal!6]| 2 & |[fill=teal!5]| 1 & |[fill=teal!15]| \textbf{2} \\
  };
  \draw[line width=0.6pt, black]
    ([xshift=-4pt,yshift=4pt]M-1-1.north west) rectangle ([xshift=4pt,yshift=-4pt]M-4-4.south east);
\end{tikzpicture}